\documentclass[a4,12pt]{article}
\usepackage{graphicx}
\newcommand{\nn}{\nonumber\\}

\newcommand{\ket}[1]{\left| #1 \right>}
\newcommand{\abs}[1]{\left| #1 \right|}

\renewcommand{\thepage}{}
\makeatletter
\@addtoreset{equation}{section}
\renewcommand{\theequation}{\thesection.\@arabic\c@equation}
\makeatother
\renewcommand{\thefootnote}{\fnsymbol{footnote}}
\makeatletter
  \def\mathcomposite{%
     \@ifstar
        {\def\@mathcomposite@option{%
            \baselineskip\z@skip\lineskiplimit-\maxdimen}%
         \@mathcomposite}%
        {\let\@mathcomposite@option\offinterlineskip
         \@mathcomposite}}
  \def\@mathcomposite{%
     \@ifnextchar[\@@mathcomposite{\@@mathcomposite[0]}}
  \def\@@mathcomposite[#1]#2#3#4{%
     #2{\mathchoice
        {\@mathcomposite@{#1}{#3}{#4}\displaystyle{1}}%
        {\@mathcomposite@{#1}{#3}{#4}\textstyle{1}}%
        {\@mathcomposite@{#1}{#3}{#4}%
         \scriptstyle\defaultscriptratio}%
        {\@mathcomposite@{#1}{#3}{#4}%
         \scriptscriptstyle\defaultscriptscriptratio}}}
  \def\@mathcomposite@#1#2#3#4#5{%
     \vcenter{\m@th\@mathcomposite@option
        \dimen@\f@size\p@\dimen@#1\dimen@\dimen@#5\dimen@
        \divide\dimen@ 18
        \edef\@mathcomposite@skipamount{\the\dimen@}%
        \ialign{\hfil$#4##$\hfil\cr
           #2\crcr
           \noalign{\vskip\@mathcomposite@skipamount}%
           #3\crcr}}}
  \makeatother
\newcommand{\lsim}{\mathcomposite{\mathrel}{<}{\sim}}
\begin{document}
\begin{titlepage}
\title{
\vspace*{-4ex}
%\hfill{\normalsize hep-th/yymm.nnnn}\\
\vspace{4ex}
\bf Level truncation analysis of exact solutions in open string field
 theory  
\vspace{5ex}}
\author{Tomohiko {\sc Takahashi}\footnote{E-mail address:
tomo@asuka.phys.nara-wu.ac.jp}\\
\vspace{2ex}\\
{\it Department of Physics, Nara Women's University, Japan}}
\date{\today}
\maketitle
\vspace{7ex}

\begin{abstract}
\normalsize
\baselineskip=19pt plus 0.2pt minus 0.1pt
We evaluate vacuum energy density of Schnabl's solution using the level
truncation calculation and the total action including interaction
terms. The level truncated solution provides vacuum
energy density expected both for tachyon vacuum and trivial pure
gauge. We discuss the role of the phantom term to reproduce
correct vacuum energy. 
\end{abstract}
\end{titlepage}

%\tableofcontents
%\vspace{1cm}

%%%%%%%%%%%%%%%%%%%%%%%%%%%%%%%%%%%%%%%%%%%%%%%%%%%%%%
\renewcommand{\thepage}{\arabic{page}}
\renewcommand{\thefootnote}{\arabic{footnote}}
\setcounter{page}{1}
\setcounter{footnote}{0}
\baselineskip=19pt plus 0.2pt minus 0.1pt
%%%%%%%%%%%%%%%%%%%%%%%%%%%%%%%%%%%%%%%%%%%%%%%%%%%%%%
%
%%%%%%%%%%%%%%%%%%%%%%%%%%%%%%%%%%%%%%%%%%%%%%%%%%%%%%%%%%%%%%
\section{Introduction}

String field theory (SFT) provides a non-pertubative framework to
analyze various string backgrounds in a unified way. Several classical
solutions have been constructed in SFT numerically and
analytically, and each of them represents the tachyon vacuum, backgrounds
with marginal deformations or rolling tachyon and so 
on~\cite{rf:SZ}--\cite{rf:tomo}.
%\cite{rf:SZ,rf:Schnabl,Ellwood:2006ba,
%Schnabl:2007az,Kiermaier:2007ba,Fuchs:2007yy, 
%Ellwood:2007xr,Kishimoto:2007bb,Kiermaier:2007vu,Kluson:2002ex,
%Kluson:2002hr,Kluson:2002gu,Kluson:2002te,Kluson:2002av,
%rf:TT1,rf:TT2,rf:KT,rf:tomo}.  
The most important progress of recent works in SFT is that Schnabl
constructed an analytic classical solution~\cite{rf:Schnabl} 
in Witten's open bosonic SFT~\cite{Witten}. The solution is represented
as
\begin{eqnarray}
 \Psi(\lambda)=\lim_{N\rightarrow \infty}
 \left[\lambda^{N+1} \psi_N-\sum_{n=0}^N 
 \lambda^{n+1}\partial_n \psi_n
 \right], \nonumber
\end{eqnarray}
where $\psi_n$ denote wedge states with certain ghost and anti-ghost
insertions, and $\lambda$ is a real parameter. It is believed that
for $\lambda=1$ the solution corresponds to the non-pertubative tachyon
vacuum and otherwise the solution should be referred to a trivial pure
gauge configuration. It is partly because
the above wedge based expression provides correct
vacuum energy density expected for the tachyon and the trivial
solutions~\cite{rf:Schnabl}.

The crucial difference between the tachyon vacuum and the trivial pure
gauge solution seems to be included in the first term of the above
expression, so-called the phantom term. Obviously, this term becomes
$\psi_\infty$ at $\lambda=1$, and if $\abs{\lambda}<1$ it is equal to
zero due to the factor $\lambda^{N+1}$\ ($N\rightarrow \infty$). 
Actually, if the first term is not involved in the solution, we can
not derive the correct vacuum energy from analytic calculation using
quadratic parts of the action~\cite{rf:Schnabl}.
Besides, it is pointed out that the first term is indispensable for the
equation of motion contracted with the solution to be
satisfied~\cite{Okawa:2006vm}\cite{Fuchs:2006hw}. In 
other words, the first term is needed to calculate the
vacuum energy using the total action with cubic terms, instead of the
quadratic action reduced by the equation of motion. 

In spite of the important effect of the phantom term, it is known that
it becomes to be ``zero'' also for the case $\lambda=1$. More precisely,
the inner product of $\psi_N$ with any Fock space state vanishes for
taking the $N\rightarrow \infty$ limit, and therefore the first term is
regarded as zero in the Fock space representation. Consequently, it is
often said that the phantom term is representative of analytic solutions
beyond the Fock space expression.
  
Interestingly, the analytic solution is regular from the viewpoint of
level truncation as pointed out in the first place~\cite{rf:Schnabl}. In fact, the
solution for $\lambda=1$ reproduces the correct vacuum energy density in
level truncation with respect to the $L_0$ operator. This energy density
was calculated only by using the quadratic action and it is never
affected by the phantom term  because the truncated solution is a state
inside the Fock space. 
Here, it is natural to ask whether the correct energy density
can be reproduced from level truncated calculation using the total
action with cubic terms, despite the crucial term is irrelevant.
To examine this question is the main motivation in this paper.

In the following, we will calculate the vacuum energy density
numerically by truncating the analytic solution and using the action
with and without the cubic terms.
We will evaluate it for all values of $\lambda$ providing a regular
solution in the level truncation calculation, although only the
$\lambda=1$ case was evaluated so far using the quadratic action. 
Finally, we will discuss the role of the phantom term to yield
the correct vacuum energy density in the last section.

%%%%%%%%%%%%%%%%%%%%%%%%%%%%%%%%%%%%%%%%%%%%%%%%%%%%%%%%%%%%%%
\section{Level truncation of the analytic solution}

The analytic solution $\Psi(\lambda)$ can be written as
\begin{eqnarray}
\label{Eq:solution2}
 \Psi(\lambda)=-\sum_{n=0}^\infty \lambda^{n+1}\partial_n\psi_n,
\end{eqnarray}
where we first take the $N\rightarrow \infty$ limit and use the fact
that $\psi_\infty=0$. Strictly speaking, this expression is correct in
the Fock space representation.
From the definition of $\psi_n$ in ref.~\cite{rf:Schnabl}, we can write
the solution explicitly as
\begin{eqnarray}
 \Psi(\lambda)&=&-\frac{1}{\pi}\sum_{n=2}^\infty
 \lambda^{n-1}\frac{d}{dn}\left\{
 U_n^\dagger\left[\frac{n}{\pi}\,{\cal B}_0^\dagger\,
   \tilde{c}\left(-\frac{\pi}{2}\frac{n-2}{n}\right)
   \tilde{c}\left(\frac{\pi}{2}\frac{n-2}{n}\right)\right.\right.\nn
&&\left.\left.
   +\tilde{c}\left(-\frac{\pi}{2}\frac{n-2}{n}\right)
   +\tilde{c}\left(\frac{\pi}{2}\frac{n-2}{n}\right)
\right]\right\}\ket{0}.
\end{eqnarray}
This expression is almost the same as that given by Schnabl except for
inclusion of $\lambda$.
After operating the ghost fields on the Fock vacuum, we find
\begin{eqnarray}
 (\tilde{c}(-x)+\tilde{c}(x))\ket{0}&=& 2\cos^2 x\,c_1
   +2\sin^2 x\,c_{-1}\ket{0} +2\cos^2 x\tan^4 x\,c_{-3}\ket{0}+\cdots\\
 \tilde{c}(-x)\tilde{c}(x)\ket{0}&=& 
   -2\cos^4 x\tan x\,c_0 c_1 \ket{0}
   -2\cos^4 x \tan^3 x(c_0c_{-1}+c_{-2}c_1)\ket{0}\cdots.
\end{eqnarray} 
The operator ${\cal B}_0^\dagger$ is expanded by negative modes of usual
anti-ghost oscillators.
The operator $U_n^\dagger$ can be expressed in the
canonically ordered form as
\begin{eqnarray}
 U_n^\dagger = \cdots e^{u_6\,L_{-6}}e^{u_4\,L_{-4}}e^{u_2\,L_{-2}}
\left(\frac{2}{n}\right)^{L_0},
\end{eqnarray}
where $u_n$ are real numbers as given in ref.~\cite{rf:Schnabl}.
These equations allows us to express the analytic solution as a
state in the Fock space. 

For example, let us expand the solution up to level 2:
\begin{eqnarray}
 \Psi(\lambda)= t\,c_1\ket{0}
  +u\,c_{-1}\ket{0}+v\,(\alpha_{-1}\cdot \alpha_{-1})c_1\ket{0}
  +w\,b_{-2}c_0c_1\ket{0}+\cdots.
\end{eqnarray}
The component fields, $t$, $u$, $v$ and $w$, are given as infinite
serieses in the following,
\begin{eqnarray}
 t(\lambda)&=& \sum_{n=2}^\infty
       \lambda^{n-1}\frac{d}{dn}
       \left[\frac{n}{\pi}\sin^2\left(\frac{\pi}{n}\right)
       \left(-1+\frac{n}{2\pi}\sin\left(\frac{2\pi}{n}\right)
       \right)\right],\\
 u(\lambda)&=& \sum_{n=2}^\infty
       \lambda^{n-1}\frac{d}{dn}
       \left[\left(\frac{4}{n\pi}-\frac{n}{\pi}\sin^2\left(
       \frac{\pi}{n}\right)\right)\left(
       -1+\frac{n}{2\pi}\sin\left(\frac{2\pi}{n}\right)
       \right)\right],\\
 v(\lambda)&=& \sum_{n=2}^\infty
       \lambda^{n-1}\frac{d}{dn}
       \left[\left(\frac{2}{3n\pi}-\frac{n}{6\pi}\right)
       \sin^2\left(\frac{\pi}{n}\right)\left(
       -1+\frac{n}{2\pi}\sin\left(\frac{2\pi}{n}\right)
       \right)\right],\\
 w(\lambda)&=& \sum_{n=2}^\infty
       \lambda^{n-1}\frac{d}{dn}
       \left[\sin^2\left(\frac{\pi}{n}\right)\left(
       \frac{8}{3n\pi}-\frac{2n}{3\pi}+\frac{n^2}{3\pi^2}
       \sin\left(\frac{2\pi}{n}\right)\right)\right].
\end{eqnarray}
All of these converge absolutely if $\abs{\lambda}\leq 1$ and
the same is true up level 10.
A component field is given by a power series as $\sum_{n=2}^\infty
\lambda^{n-1}a_n$ if it is non-zero. We can easily check that the radius
of convergence is 
1 for these component fields up to level 10. For $\abs{\lambda}=1$, we
expand as $a_n/a_{n+1}=1+h/n+{\rm O}(1/n^2)$ and we can find $h>1$ up to
level 10. Hence, the series is convergent for $\abs{\lambda}\leq1$.

The expression (\ref{Eq:solution2}) satisfies the equation
of motion for arbitrary $\lambda$, that is proved only by using the
identity of $\psi_n$ irrelevant to $\lambda$ \cite{rf:Schnabl}.
It is not clear for what range of $\lambda$ the solution should be
defined. However, $\lambda$ must take the value between $-1$ and $1$ if
the solution has a well-defined Fock space expression.

It is difficult to derive analytic expressions for these serieses.
If we expand coefficients in the series in powers of $1/n$, only terms
$1/n^4$, $1/n^6$, $1/n^8$, $\cdots$ appear in it. Therefore, 
we can sum up the series numerically with extreme precision as mentioned
in ref.~\cite{rf:Schnabl}. 
These fields obey a symmetry generated by $K_1^{matter}$ for any
$\lambda$ as in case of $\lambda=1$ \cite{rf:Schnabl}. Using this
symmetry, we can check numerically whether the calculated result is 
correct or not. 
The resulting plots of the above fields are depicted in Fig.~1. These
values at $\lambda=1$ 
coincides with those of earlier results in ref.~\cite{rf:Schnabl}. Each
curve for component fields smoothly varies from zero at $\lambda=0$. It
has no discontinuity even at $\lambda=1$, despite the vacuum energy
should fall down from zero to the minus energy at $\lambda=1$.
\begin{figure}[h]
\centerline{\includegraphics[width=9cm]{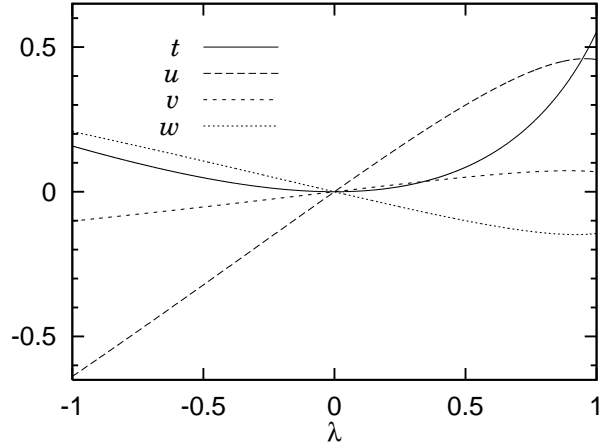}}
\caption{Component fields up to level 2. Each curve consists of two
 thousand plots with lines.}
\end{figure}

While the $\lambda\neq 1$ case is expected to be trivial pure gauge, the
$\lambda=-1$ case is exceptional because the solution for $\lambda=-1$
satisfies the symmetry
\begin{eqnarray}
 (-1)^{{\cal L}_0} {\cal L}_0 \Psi={\cal L}_0\Psi.
\nonumber
\end{eqnarray}
It is the same symmetry satisfied by the tachyon vacuum solution and
therefore the solution in that case may be regarded as a non-trivial
vacuum~\cite{rf:Schnabl}. However, at the $\lambda=-1$, all component
fields are continuous similar to those of $\lambda=1$.
\begin{figure}[h]
\centerline{\includegraphics[width=9cm]{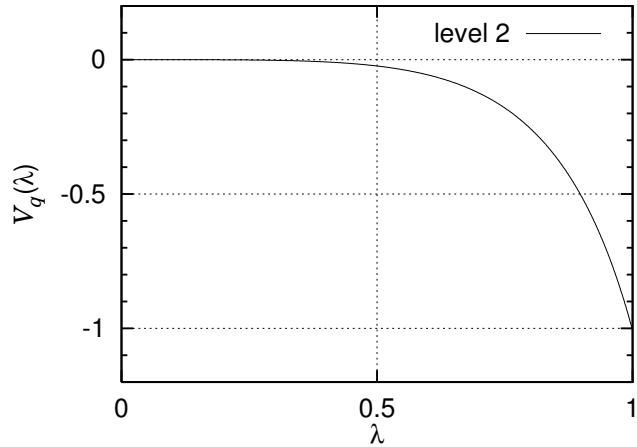}}
\caption{The energy density for level 2.
It is calculated only with the quadratic terms of the action.}
\end{figure}

\begin{figure}[h]
\centerline{\includegraphics[width=9cm]{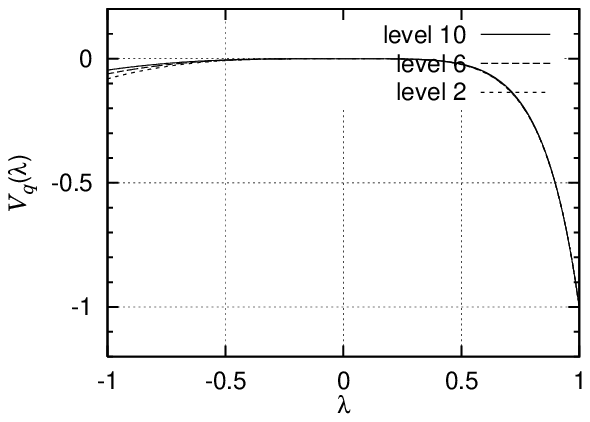}}
\caption{The energy densities up to level 2, 6, 10 which
are calculated only with the quadratic terms of the action.
Each line is drawn as two thousand points with lines.}
\end{figure}

\begin{figure}[h]
\centerline{\includegraphics[width=9cm]{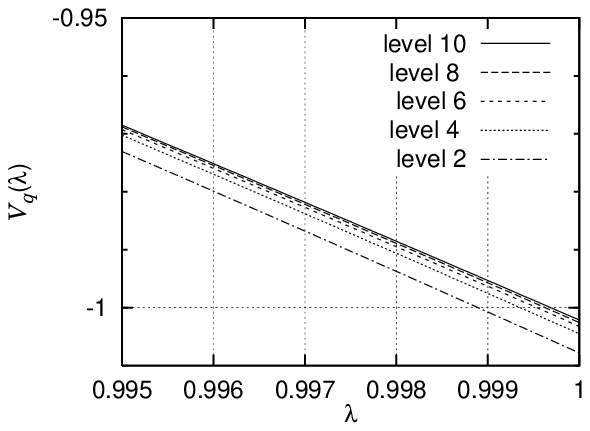}}
\caption{The energy densities up to level 10 which
are calculated only with the quadratic terms of the action.
Each line is drawn as one thousand points with lines.}
\end{figure}

Now, let us compute the vacuum energy using the action with the
quadratic terms only. Using the equation of motion, the vacuum energy
density is given by
\begin{eqnarray}
 V_{q}(\lambda)=\frac{\pi^2}{3}\left<\Psi(\lambda),\,
  Q_B\Psi(\lambda)\right>, 
\end{eqnarray}
where it is normalized as to be minus one for the tachyon vacuum.
Substituting the truncated solution into it, we can calculate
the energy density as a function of $\lambda$. For level 2 and $0\leq
\lambda \leq 1$, we make 
a graph of the vacuum energy density in Fig.~2. 
As given in ref.~\cite{rf:Schnabl}, the energy density at $\lambda=1$ is
good agreement with the correct density even at level 2.
For $0<\lambda \lsim 0.3$, the energy density is almost zero and it is
well-behaved as a pure gauge solution. We can not understand this zero
energy density trivially by the values of component fields in Fig.~1.
This good property can be regarded as a result of cancellation of each
contribution of all component fields.

We consider higher level approximation for full range of $\lambda$.
We have computed component fields up to level 10. We display the result
in Fig.~3. 
Around $\lambda=0$, we have almost zero energy density for all level.
For $\lambda\sim -1$, the energy density approaches zero as the
truncation level is increased. This result is consistent with the
expectation that the solution is a trivial pure gauge solution for
$-1\leq \lambda <1$.
Around nearby $\lambda=1$, we can not distinguish each curve from the
others. So, we enlarge the resulting plots for higher levels around
$\lambda\sim 1$ in Fig.~4. 
We find that the energy density approaches slowly but gradually to the
correct value as the approximation level is increased.
Consequently, although the plots may approach a critical curve for
higher levels, the result is consistent with the expectation that, if
the truncation level goes to infinity,  the plots approach to the step
function,
\begin{eqnarray}
 f(\lambda)=\left\{
\begin{array}{ll}
 0 & (-1\leq \lambda <1)\\
 -1 & (\lambda =1).
\end{array}\right.
\end{eqnarray}

Next, let us compute the energy density using the total action including
cubic terms, which is given by
\begin{eqnarray}
 V_{f}(\lambda)=\frac{\pi^2}{2}\left<\Psi(\lambda),\,
  Q_B\Psi(\lambda)\right>+\frac{\pi^2}{3}
 \left<\Psi(\lambda),\,\Psi(\lambda)*\Psi(\lambda))
 \right>,
\end{eqnarray}
where we have used the same normalization as before. 

For $\lambda=1$,
the resulting energy density is summarized in the following table.
\begin{table}[h]
\begin{center}
\begin{tabular}{|c|c|c|c|c|c|c|c|}
 \hline
&$L=0$ & $L=2$ & $L=4$ & $L=6$& $L=8$& $L=10$\\
 \hline
 \hline
$(L,\,2L)$ & -0.577920 & -1.081077 & -1.054081 & -1.036779 &
   -1.025645 & -1.018552 \\
 \hline
$(L,\,3L)$ & -0.577920 & -1.065177 & -1.047979 & -1.032868 &
   -1.023261 & --- \\
 \hline
 \hline
quad. terms & -1.007766 & -1.007815 & -1.004499 & -1.003217 & -1.002556 &
   -1.002130  \\
 \hline
\end{tabular}
\end{center}
\caption{Energy density calculated by the full action.
For comparison, energy density calculated by the quadratic action is
 listed in the last row.}
\end{table}
For level zero, the energy density is $-0.57\cdots$ and it is about
one-half of the correct density. But, it is comparable to the level zero
result in Siegel gauge, $-0.68\cdots$. The both results of $(L, 2L)$ 
and $(L, 3L)$ are almost the same.
The level 6 and 8 results of $(L,3L)$ are closer to $-1$ than that of
$(L,2L)$. Up to level 10, the energy density agrees with the correct
value to $10^{-1}$, but that is 
worse than the result calculated by the quadratic action. However, we
can find that the resulting energy approaches to the expected value -1
as the truncation level is increased.

We proceed to consider the full range of $\lambda$. In Fig.~5 and 6, we
display the energy density evaluated by the truncated actions of
$(L, 2L)$ and $(L, 3L)$, respectively. Around nearby $\lambda=0$, the energy
density is almost equal to zero. Around $\lambda=-1$, the energy density
approaches to zero as the truncation level is increased, but
from the positive energy region as contrasted to the calculation by the
quadratic action. Even in the case using the total action, we can find
that the resulting plots gradually become the step function as the
level is increased.

\begin{figure}[h]
\centerline{\includegraphics[width=10cm]{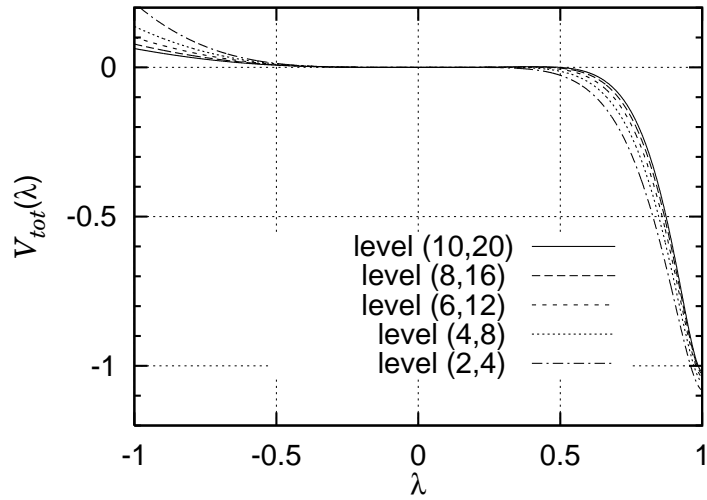}}
\caption{Energy density evaluated by the $(L,2L)$ truncated action.}
\end{figure}

\begin{figure}[h]
\centerline{\includegraphics[width=10cm]{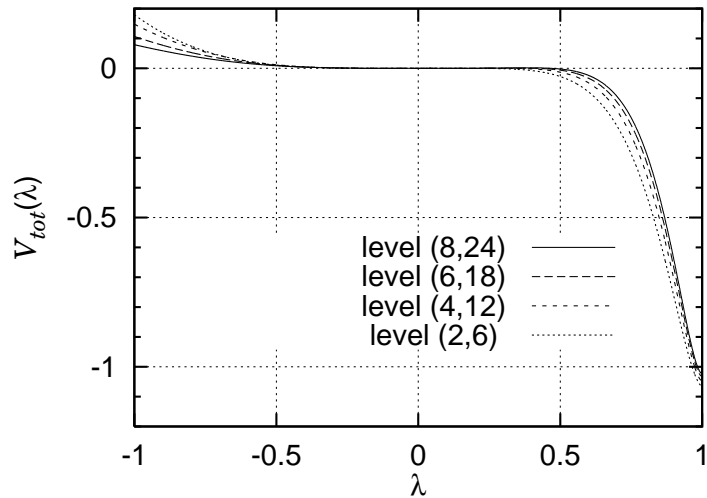}}
\caption{Energy density evaluated by the $(L,3L)$ truncated action.}
\end{figure}

\newpage
%%%%%%%%%%%%%%%%%%%%%%%%%%%%%%%%%%%%%%%%%%%%%%%%%%%%%%%%%%%%%%
\section{Discussions}

We calculated the vacuum energy density for the analytic classical
solution constructed by Schnabl using its Fock space expression.
We found that, as the truncation level is increased, the resulting plots
approach to the step function for $\abs{\lambda}\leq 1$. The result is
consistent with the fact that the solution for $\lambda=1$ is the
tachyon vacuum solution and otherwise it corresponds to a trivial pure
gauge solution. In particular, our calculation suggests that the
solution for $\lambda=-1$ is also trivial, although it possesses
the same symmetry as the tachyon vacuum solution. Consequently, the
analytic solution is well-behaved for $\abs{\lambda}\leq 1$ from the
point of view of level truncation.

Our analysis was based on level truncation calculation and therefore the
phantom term does not contribute to the vacuum energy density.
Our result suggests that the phantom term is not indispensable to
reproduce the correct vacuum energy, although the phantom term is an 
important ingredient to evaluate the vacuum  energy analytically.

It is no wonder that the phantom term plays a whole different
role in each expression of the solution. Because, a string field
is given as a state in the Hilbert space with an indefinite metric,
namely string field theory does not possess a positive definite norm. 
Actually, the correct vacuum energy was reproduced as a result of
cancellation between positive and negative infinite energy. This fact
can be found most clearly in the solution expanded in ${\cal L}_0$
eigen-states. Using the solution truncated with respect to the ${\cal 
L}_0$ level, we find that the vacuum energy density is not convergent as
the truncation level is increased. But, surprisingly, the Pad\'e
approximation to the divergent series can reproduce correct vacuum
energy both for $\lambda=1$~\cite{rf:Schnabl} and
$\lambda\neq 1$~\cite{rf:ikishimo}. This result of the 
${\cal L}_0$ truncation indicates that the 
vacuum energy density for the solution is given as a conditionally
convergent series.

Hence, to define the analytic solution, an important point
is how to regularize it in SFT based on the indefinite
metric. In the wedge based expression, the integer $N$ seems to be a
kind of regularization parameter. The phantom term is important only if
we regularize the solution in terms of $N$.
Our results suggest that the truncation level $L$ can be regarded as a
good regularization parameter as well as $N$. If that is the case, we
will find that the vacuum energy from the truncated solution agrees with
the correct value as the $L$ goes to infinity, that is
\begin{eqnarray}
 \lim_{\epsilon\rightarrow 0} 
\frac{\pi^2}{3}\left<e^{-\epsilon L_0}\Psi(\lambda),\,Q_B
e^{-\epsilon L_0}\Psi(\lambda)\right>=
\left\{\begin{array}{ll}
 -1&\ \ \ (\lambda=1)\\
 0&\ \ \ (-1\leq \lambda <1).
\end{array}
\right.
\end{eqnarray}
We expect a similar behavior for the vacuum energy including
the cubic terms. In any case, the phantom term has no effect on the
conjectural equation. 

The parameters $N$ and $L$ seem to provide a sort of ``correct''
regularization. Eventually, a crucial issue is how we can regularize the
solution or the theory ``correctly''.
It is well-known that symmetry is a key role to regularize a quantum field
theory of gauge fields, which is formulated in the framework of the
indefinite-metric theory. 
In contrast, we still lack the criterion for ``correct'' regularization
in string field theory.
For example, the vacuum energy for the identity-based solution is given
as an indefinite quantity~\cite{rf:tomo}~\cite{Igarashi:2005wh}.
We hope that the Schnabl's solution will help to
find a good way to regularize string field theory in order to search
non-pertubative vacua further.

%%%%%%%%%%%%%%%%%%%%%%%%%%%%%%%%%%%%%%%%%%%%%%%%%%%%%%%%%%%%%%
\section*{Acknowledgments}
The author would like to thank T.~Kawano, T.~Kugo and especially
I.~Kishimoto for useful discussions. The author thanks also the Yukawa
Institute for Theoretical Physics at Kyoto University, and 
the Institute of Physical and Chemical Research.
Discussions during the YITP workshop YITP-W-07-05 on ``Quantum Field
Theory 2007''  and the RIKEN Symposium ``String Field Theory 07'' 
were useful to complete this work.
This work was supported in part by a Grant-in-Aid for Young Scientists
(B) (\#18740152) from the Ministry of Education, Culture, Sports,
Science and Technology of Japan.

%%%%%%%%%%%%%%%%%%%%%%%%%%%%%%%%%%%%%%%%%%%%%%%%%%%%%%%%%%%%%%

%%%%%%%%%%%%%%%%%%%%%%%%%%%%%%%%%%%%%%%%%%%%%%%%%%%%%
%%%%%%%%%%%%%%%%%%%%%%%%%%%%%%%%%%%%%%%%%%%%%%%%%%%%%
\end{document}